\documentstyle[preprint,aps,prl]{revtex} 
\tightenlines
\begin{document}
\newcommand{\sinf}{\raisebox{-.7ex}{$\stackrel{<}{\sim}$}}
\newcommand{\ssup}{\raisebox{-.7ex}{$\stackrel{>}{\sim}$}}

\ifx\undefined\psfig\def\psfig#1{    }\else\fi     \ifpreprintsty\else
\twocolumn[\hsize\textwidth
\columnwidth\hsize\csname@twocolumnfalse\endcsname       \fi    \draft
\preprint{  }  \title  {Continuum elasticity theory of edge 
excitations in a two-dimensional electron liquid with finite range 
interactions}  
\author  {Irene D'Amico and Giovanni Vignale}
   \address{Department  of Physics,   University   of
Missouri, Columbia, Missouri 65211} \date{\today} \maketitle

\begin{abstract}

We make   use  of continuum    elasticity theory  to  investigate  the
collective  modes that propagate along the  edge  of a two-dimensional
electron liquid or crystal in a magnetic field.   An exact solution of
the  equations  of motion is  obtained  with the following simplifying
assumptions: (i) The system  is {\it macroscopically} homogeneous  and
isotropic  in   the half-plane   delimited   by   the  edge  (ii)  The
electron-electron interaction is of  finite range due to  screening by
external electrodes  (iii) The  system  is nearly  incompressible.  At
sufficiently  small wave vector  $q$  we  find a universal  dispersion
curve $\omega   \sim q$ independent  of the  shear modulus. At larger
wave vectors the dispersion can change  its form in a manner dependent
on  the comparison  of  various length  scales.  We obtain  analytical
formulas   for the  dispersion  and  damping of  the modes  in various
physical regimes. 
\end{abstract} 
 \ifpreprintsty\else\vskip1pc]\fi \narrowtext 
\section{Introduction} 
The dynamical  behavior of the edge  of a two dimensional electron gas
(2DEG)  in  a  strong magnetic  field is receiving considerable
attention.    The   interest has been   spurred   by recent experiments
\cite{Grayson,Chang},  in  which an unusual  tunneling current-voltage
relation of the  form $I =  V^{1/\nu_0}$, where $V$  is the bias voltage
and  $\nu_0$   is the bulk  filling  factor,  has  been observed. Such a
distinctly   non-Fermi-liquid-like  behavior  can     be      
explained
\cite{Conti1,Conti2,Zulicke} on the assumption that there is
only one branch of  collective edge wave excitations (the so-called
charged mode), 
and that the tunneling
electron is initially accommodated as a coherent superposition of such
waves spreading in time according to hydrodynamic equations of motion. 

Unfortunately,   a complete  theory   of edge excitations  is not  yet
available.  For example,  there is considerable  controversy about the
existence of additional ``neutral modes'', and whether they contribute
to the tunneling characteristics or not \cite{Shytov,Lee,Lopez}. 

Even the more conventional charged mode has not been fully analyzed yet. 
The two most successful theories so far \cite{Volkov,Aleiner} focus
exclusively on the long-range part of the Coulomb interaction and should
therefore be viewed as macroscopic versions of the random phase
approximation (RPA).  A complete description of the long wave dynamics
requires, however, the inclusion of short range forces that arise from the
relative motion of adjacent parts of the fluid.  These can be pressure
forces arising from the bulk modulus of the quantum mechanical fluid, as
well as shear forces arising from more subtle positional correlations.
Shear forces are, of course, essential in the crystalline phase, but they
are also present in the liquid phase at non zero frequency \cite{Conti98}. 
In addition, there are viscous forces - completely ignored in the RPA -
that cause damping of the collective modes. 

	These  post-RPA effects are well  established in the theory of
the dynamics  of the  uniform  electron gas,  where  they  are usually
described in    terms     of  ``local    field     corrections''  (see
\cite{Conti98}).  For purely longitudinal bulk modes these corrections
are not too important  as the direct  Coulomb interaction controls the
physics of long  wavelength longitudinal fluctuations.  The  situation
can be quite  different for edge modes, or   even for bulk modes  in a
magnetic field, because  in  these cases longitudinal   and transverse
channels   are  strongly   coupled,  and   one  cannot    talk of purely
longitudinal or   transverse modes.  Indeed,    we have recently shown
\cite{Conti2} that the collective dynamics of a {\it uniform} electron
gas, in the limit of infinite magnetic field, must necessarily include
a shear force term, otherwise  the frequencies of the collective modes
vanish.   This led  us  to the description of  the  electron  gas as a
continuum elastic medium, characterized by elastic constants (the bulk
and the  shear modulus) which  control the dynamical  response, and by
viscosity coefficients, which control dissipation. 

Encouraged   by  the   qualitative  success of  the  continuum 
elasticity
approach to the dynamics of the uniform electron gas, in this paper we
present its application to the problem of edge  dynamics in a magnetic
field. 

On a macroscopic scale, our system is modeled as a uniform continuum in
the half-plane delimited by the edge.  The external potential, that
confines the electrons to the half-plane, is constant in the bulk of the
system, and rises sharply at the edge.  A great mathematical
simplification follows from the assumption that the electron-electron
interaction is of finite ( microscopic) range. This assumption is
justified for structures in which the long-range Coulomb interaction is
screened by metallic electrodes.  The short range of the interaction
results in equations of motion for the elastic displacement field that
are linear differential equations, as opposed to integral equations in the
approach of \cite{Volkov,Aleiner}.  

The distinctive feature of our
theory is that the presence of the edge is taken into account via {\it
boundary conditions}. In an
infinite system, the solutions
of the equations of motion are purely longitudinal or transverse plane
waves characterized by a {\it real} wave vector.
  The presence of the edge allows the existence of solutions with an {\it
imaginary} wave vector perpendicular to the edge, which therefore vanish
exponentially as one moves into the bulk.  It is evident that these
``bound" solutions exist independently of the nature of the bulk, in
particular, irrespective of whether the bulk is incompressible or not. In
addition, these solutions must have the property that the elastic stress
vanish at the edge.  Because the  boundary conditions are linear, the
problem can be solved exactly by elementary techniques, and analytic
results for the dispersion and damping of the collective modes can be
obtained.  (The qualitative changes brought about by the long range of the
Coulomb interaction will be discussed in the Appendix). A particularly
elegant solution can be obtained for nearly incompressible systems, in
which a large electrostatic charging energy strongly opposes density
fluctuations. 

Our main result is that, at  sufficiently long
wavelengths and high magnetic field there is only {\it  one}
low-frequency edge mode, whose dispersion 
depends  on the strengths of  the confining electric and magnetic fields,
but   not on the  bulk and   shear moduli of   the
system. The nature of this solution does not change when
the system  has  crystalline order.    At
shorter wavelengths, the dispersion  can change its  form in a  manner
dependent on the comparison  of various length scales.  In particular,
we point out the possibility of  a crossover from a linear dispersion,
controlled by the confining electric field, to a quadratic dispersion,
controlled by  the  shear modulus of  the  electronic  system,  if the
latter is sufficiently large. 

   Our results support the idea that the dynamics of a sharp edge in a
2DEG is completely dominated by a {\it single} charged mode, with any
additional  structure associated with  peculiar characteristics of the
system (such as the fractional quantum Hall effect) becoming irrelevant
at long wavelengths. 

\section{The model} We consider a two dimensional electron fluid on a
uniform background of  positive charges, in  a constant magnetic field
${\bf B}=B{\hat {\bf z}}$. The  system extends indefinitely for $x<0 $
while presenting a sharp edge at $x \sim 0$, parallel to the $y$ axis.
Since  the   width  $a$ of   the    edge is  much  smaller   than  the
characteristic wavelength of  the modes under consideration,  we model
the edge   setting   $a  \simeq 0$.  The    {\it   mass} density,   at
equilibrium, is a constant $\rho_0$ for $x<0$, and vanishes for $x>0$.
Consistent with this  assumption, the external electrostatic potential
$V_{ext}(x)$, created  by  the background of  positive  charges and by
nearby gates, is taken to be constant for $x<0$ and to rise sharply at
$x = 0$ with a derivative 
\begin {equation} \label{defg} g \equiv {e \over m} \left ( {d
V_{ext}  \over dx} \right )_{x=0}.  \end {equation} 
Here $-e$ is the {\it effective} electron charge, i.e., the bare charge 
divided by the square root of the static dielectric constant of the host
 semiconductor,  and $m$ is the {\it  effective} mass of the electron 
in the host semiconductor.   The  quantity   $g$  is  the
acceleration imparted by the external potential  to an electron at the
edge.  The effect of this acceleration on the collective edge modes is
analogous to   the effect  of  gravity on  surface  waves in a liquid -
hence the notation.

To describe  the dynamics of the  system we introduce the displacement
field  ${\bf u}(x,y,t)$ of  the infinitesimal volume  element at point
$(x,y)$ from its equilibrium   position.  This obeys  the   linearized
equation of  motion  of  continuum elasticity  theory  \cite{Landau7}:
\begin{equation} 
\label{equationofmotion1} \frac{\partial^2{u}_i}{\partial t^2}={1 \over
\rho_0}\sum_j \partial_j  \sigma_{ij}+   \omega_c(\hat {\bf  z}\times{
\partial {\bf u}  \over \partial t})_i  \end{equation} where $i,j$ are
Cartesian indices,  $\partial_i$   is the derivative  with  respect to
${r}_i$, $\omega_c = eB/mc$ is the cyclotron frequency, and 
\begin{equation} \label {sigma} \sigma_{ij}=K\nabla\cdot{\bf
u}\delta_{ij}+    \mu(\partial_iu_j+\partial_ju_i     -\nabla\cdot{\bf
u}\delta_{ij}) \end{equation} is the  elastic stress tensor  with bulk
modulus $K$ and shear modulus $\mu$. The  first term on the right hand
side of eq. (\ref{equationofmotion1}) is  the force exerted  on the 
volume   element by the surrounding
medium,  the  second term  is  the Lorenz  force.    In writing these
equations, we have assumed that the electron-electron
interaction $v({\bf r} - {\bf r'})$ is  of finite range in space. This
means that  $v({\bf q}=0) \equiv \int  v({\bf r})d  {\bf r}$ is finite
and  is included as part  of the bulk modulus:  \begin  {equation} K =
v({\bf  q} = 0) n_0^2  + \tilde K, \end
{equation} where $\tilde K$ is the ``proper" contribution arising from
the kinetic and exchange-correlation energy and $n_0 \equiv \rho_0/m$ is
the particle density. This procedure is, of course,
only justified when the spatial variation of the density is small
over distances of the order of the range of the interaction $d$, so that $qd 
\ll 1$, where $q$ is the wave vector. We must also have $qa \ll 1$ for the 
``sharp edge'' description to be valid. 
Clearly, both conditions are satisfied at sufficiently long wavelengths.


Substituting eq.~(\ref{sigma}) into   eq.~(\ref{equationofmotion1}) we
obtain the standard form of the equation of motion 
\begin{equation}
\frac{\partial^2{\bf  u}}{\partial t^2}=C^2\nabla(\nabla\cdot{\bf u})+
C_t^2\nabla^2{\bf u}+\omega_c{\hat {\bf z}}\times { \bf \dot{u}}, 
\label{equationofmotion}
\end{equation}
where 
\begin{equation}
{C}_t^2={\mu \over \rho_0} , \label{C_t} 
\end{equation}
is the square of  the transverse sound velocity  in the absence of the
magnetic field and 
\begin{equation}
C^2={K \over \rho_0}=C_l^2-C_t^2 
\end{equation} 
where $C_l^2$ is the square of the  longitudinal sound velocity in the
absence of a magnetic  field.  As stated in  the introduction, we 
assume that the system is nearly incompressible, in the sense that the
charging
energy $v(q=0)n_0$ gives the dominant contribution to the bulk modulus,
and is much larger than any other energy scale in the problem, such as
$\mu/n_0$ or  $\hbar \omega_c$. This
implies that  $C_l  \gg C_t$ so  that the difference between $C^2$ and
$C_l^2$ can and will be ignored in the following. 
We emphasize that this assumption (which is expected to be reasonable for
an electrically charged system) simplifies the calculations,  but is not
otherwise essential.

To  complete  the  definition of  the  model we  must  now specify the
boundary conditions on the solution of eq.~(\ref{equationofmotion}). 
 In elasticity theory,  the normal form  of the boundary condition at a
{\it free}  surface is (\cite{Landau7})  \begin {equation} \label{bc0}
\sum_j  \sigma_{ij}{\Big\vert}_{edge} \hat  n_j = 0,   \end {equation} 
where
$i=x$ or $y$, $\hat n$ ($  = \hat x$ in this  case) is the unit vector
perpendicular  to the edge,  and the subscript  ``edge" means that the
quantity on  the left hand side must  be evaluated  at the position of
the {\it moving} edge, that is,  at the point of coordinates
$(u_x(0,y),y)$.
The physical significance of this condition is that there is no matter
beyond the edge to exert a force on the system. 

This  boundary  condition must be  slightly  modified here because the
external potential produces a stress in  the edge region even when the
system  is in equilibrium. Let us   denote by $\sigma^{(0)}_{ij}$ this
equilibrium stress.  Then the free boundary conditions take the form 
\begin {equation} \label{bc00}
\sum_j [\sigma^{(0)}_{ij}(u_x,y) + \sigma_{ij}(u_x,y)] \hat n_j = 0. 
\end {equation}

From the symmetry of the problem one sees that only the $xx$ component
of the equilibrium stress tensor is non zero, and from the equilibrium
condition $d \sigma^{(0)}_{xx}(x)/dx +e (\rho_0/m) dV_{ext}(x)/dx = 0$
one sees that 
\begin{equation} \label{equilibrium}
\left (\frac{d\sigma_{xx}^{(0)}}{dx} \right )_{x=0}=g\rho_0, 
\end{equation}
with  $g$   defined  in  eq.~(\ref{defg}).  Finally   one is free to 
choose
$\sigma^{(0)}_{xx}(x=0,y) =  0$ so as to  satisfy the free boundary
conditions at equilibrium. 

Expanding eq.~(\ref{bc00}) to first order in  ${\bf u}$, and making use
of  eq.~(\ref{equilibrium}), we finally obtain  explicit forms for the
two boundary conditions: 
\begin{eqnarray}
&  &   C_l^2\nabla\cdot{\bf   u}(0,y)+{C}_t^2(\frac{\partial u_x(0,y)}
{\partial        x}-        \frac{\partial          u_y(0,y)}{\partial
y})+gu_x(0,y)=0\label{bc1}\\  &  &  \frac{\partial  u_y(0,y)}{\partial
x}+\frac{\partial u_x(0,y)}{\partial y} =0.\label{bc2} 
\end{eqnarray}

Together with the condition that  the displacement field vanish for $x
\to     - \infty$,     eqs.~(\ref{equationofmotion})  and
 (\ref{bc1})- (\ref{bc2})
completely define the mathematical problem under consideration.

\section{Edge Waves in a Magnetic Field} 
Ignoring at first the  boundary conditions it is easy  to see that the
solutions of   eq.~(\ref{equationofmotion})  that vanish  for $x  \to
-\infty$ can be chosen to have the form 
\begin{equation}
{\bf u}(x,y,t)={\bf{\tilde u}}
e^{iqy+\lambda x-i\omega t}
\label{ans}, 
\end{equation}
where $q$ is a real wave vector parallel to the edge and $\lambda$ is,
in general,  a complex  number, whose real  part  must be  positive in
order to ensure decay in the interior of the system. 

Taking  the divergence  and  the  curl  of  the  equations of   motion
(\ref{equationofmotion})   and  making  use  of  eq.~(\ref{ans}),  we
obtain: 
\begin{eqnarray}
&    &     [-\omega^2-C_l^2(\lambda^2-q^2)]\nabla\cdot{\bf
u}-i\omega\omega_c (\nabla\times {\bf u})_z=0 \label{eqmo1}\\ &
&i\omega\omega_c\nabla\cdot{\bf u}+
[-\omega^2-{C}_t^2(\lambda^2-q^2)]  (\nabla\times   {\bf
u})_z= 0. 
\label{eqmo2}
\end{eqnarray}

These two  linear homogeneous equations are compatible  if and only if
the square of the wavevector $\lambda^2$ has one of the two values 
\begin {equation}
\label{exactlambdas}
\lambda^2 =q^2   -{\omega^2    (C_t^2+C_l^2)  \over  2     C_t^2C_l^2}
\pm\sqrt{\left (   {\omega^2 (C_t^2+C_l^2) \over 2  C_t^2C_l^2} \right
)^2 - {\omega^2 (\omega^2 - \omega_c^2) \over C_l^2C_t^2}}. 
\end {equation}
In the limit $C_l \gg C_t$ the two solutions for $\lambda^2$ (which we
denote $\lambda_l^2$ and $\lambda_t^2$) take the simple form 
\begin{eqnarray}
\lambda_l^2&=&q^2+\frac{\omega_c^2-\omega^2}{C_l^2}     \label{lam+}\\
\lambda_t^2&=&q^2-\frac{\omega^2}{{C}_t^2}-
\frac{\omega_c^2}{C_l^2}\label{lam-} 
\end{eqnarray}

With  a little algebra, it is  possible to calculate the corresponding
eigenfunctions in the limit $C_l\to \infty$: 
\begin{equation}
{\bf  u}_l \propto [ \hat{\bf   x}+i sgn(q) {\hat{\bf y}}]
e^{iqy+\vert q \vert x-i\omega t} \label{ul}\end
{equation} 
\begin {equation}
{\bf    u}_t \propto  [{    \hat{\bf x}}  +i    sgn(q)\sqrt{1  -
\omega^2/C_t^2 q^2}{\hat{\bf y}}]
e^{iqy+\vert q \vert \sqrt{1 - \omega^2/C_t^2q^2} x-i\omega t}.\end{equation} 

Let us now  turn to the  problem of satisfying the boundary conditions
~(\ref{bc2})  at  $x=0$.   This can   be done by  forming  a  suitable
superposition of the two independent solutions  ${\bf u_l}$ and
${\bf  u_t}$, namely 
\begin {equation}\label{superposition}
{\bf u} = a{\bf  u_l} + b{\bf  u_t} 
\end {equation} where $a$ and $b$ are complex coefficients.
In order to implement the  boundary conditions of eq. ~(\ref{bc2})  we
must  calculate the limit for $C_l  \to \infty$ of the products $C_l^2
{\bf  \nabla \cdot  u}_{l(t)}$.   To accomplish this, we need to
refine  our   calculation of the  eigenfunctions  ${\bf  u}_{l(t)}$ by
including    terms   of   order   $1/C_l^2$.    This    can   be  done
straightforwardly,   with the help      of  eqs.   (\ref{lam+})    and
(\ref{lam-}), and the result is 
\begin {equation}
lim_{C_l  \to \infty}  C_l^2 {\bf  \nabla \cdot   u}_l(0,y) = {\omega
[\omega_c-\omega sgn(q)] \over q} e^{iqy}\label {cnablal} \end {equation}
and 
\begin {equation}
lim_{C_l \to \infty} C_l^2  {\bf  \nabla \cdot  u}_t(0,y) =  {\omega
\omega_c \over q}e^{iqy} \label {cnablat}. \end {equation} 

Substituting eqs.~(\ref{cnablal}) and (\ref{cnablat}) in the  boundary  
conditions ~(\ref{bc1})-(\ref{bc2}) yields the  following   set of  linear
homogeneous equation for the coefficients $a$ and $b$ 
\begin {eqnarray} \label {bcq}
& & [g + {\omega\omega_c - \omega^2 sgn(q) \over q}  +2 C_t^2 \vert q
\vert ]a + [g+ {\omega \omega_c \over   q}  + 2  C_t^2   \sqrt{1 -
\omega^2/C_t^2  q^2}  \vert q  \vert] b =0 \nonumber  \\ & &   2a  + [2 -
\omega^2/C_t^2 q^2]b = 0. 
\label {bc3} \end{eqnarray}

These  two  equations  are compatible  if  and only  if  the frequency
$\omega$ satisfies the algebraic equation 
\begin{equation}
(2-{\xi}^2)^2-{\xi}^2({\xi}{       Z}+{X})          =4\sqrt{1-{\xi}^2}
\label{equationfordispersion} 
\end{equation}
where
 
\begin{eqnarray} \label{parameters}
{X} &=&\frac{g}{|q|{C}_t^2} \\
{Z}&=&\frac{\omega_c}{|q|{C}_t} \\ 
{\xi}&=&\frac{\omega}{q{C}_t}. 
\end{eqnarray}

The complete solution for the displacement field is 
\begin {eqnarray}
{\bf u}(x,y,t) & =& \{ [ (2 - \xi^2)e^{\vert q \vert x} - 2 e^{\vert q
\vert \sqrt{1 - \xi^2}x}]{\bf \hat x}\nonumber \\ & & + i sgn(q) [(2 -
\xi^2)e^{\vert q \vert x} - 2 \sqrt  {1-\xi^2}e^{\vert q \vert \sqrt{1
- \xi^2}x}]{\bf \hat y} \} e^{i(qy-\omega t)}. 
\label {eigenfunction} \end {eqnarray}

Equations ~(\ref{equationfordispersion})    and ~(\ref{eigenfunction})
are  the central   results    of  this  paper.   In  particular,   eq.
~(\ref{equationfordispersion})
is  the generalization of  the classical  equation for the
dispersion  of surface  waves  in an elastic   medium -- the so called
``Rayleigh waves''  (see ref.  \cite{Landau7}) -- to which it  reduces 
when the  dimensionless parameters  $X$ and $Z$, measuring
the strength of the external  electric and magnetic fields relative to
the shear elastic forces, are equal to zero. 
 
The solutions of eq.~(\ref{equationfordispersion}) will
be  discussed  in      the  next Section     for  various   physical
regimes. Before  doing  that, however,  it is  necessary  to clarify a
 delicate point which arises when one attempts to take the
limit $C_t \to 0$ of the above theory.  Physically this corresponds to
the very relevant  case of a genuine  liquid system, which is expected
to have a  vanishing shear modulus  at  low frequency.  It is  evident
from  eq.~(\ref{eigenfunction}) that  this   limit  is  singular:  the
variable $\xi$ tends to infinity, implying  that only the ${\bf 
u}_l$ component of the  solution  (that is,  the part proportional  to
$\xi^2$ in   eq.~(\ref{eigenfunction}))  survives.  Nevertheless,  the
presence  of a rapidly oscillating   component  of the solution   with
wavevector $\lambda \sim  i q \vert  \xi \vert$, which does not vanish
for  $x  \to - \infty$,  is  disturbing.  This difficulty becomes more
evident  when one tries to solve  the equation for  the dispersion: in
the limit $C_t \to  0$, $\xi \to \infty$, the  argument of  the square
root lies on the negative real axis,  where the presence of the branch
cut prevents us from finding a solution. 

The resolution of these  difficulties   lies in the following    physical
considerations. Every system, liquid or solid, has a non vanishing {\it
shear viscosity} $\eta$   at  finite frequency. The   shear  viscosity
contributes   an    additional  term  to   the   stress    tensor (see
\cite{Landau6}), which can be accommodated within our formalism simply
through   the replacement    $\mu  \to \mu    -  i  \omega   \eta$ or,
equivalently, \begin {equation} C_t^2 \to \tilde  C_t^2 \equiv C_t^2 -
i \omega \nu, 
\label{viscosity} \end {equation} where $\nu \equiv \eta /\rho_0$ is the
so-called kinematic viscosity \cite{Landau6}. In a solid $C_t$ remains
finite for $\omega \to 0$ and therefore $\nu$ can be safely neglected.
In a liquid, however, $ C_t$ tends to  $0$ faster than $\omega$, while
$\nu$ remains finite : the low  frequency limit is therefore dominated
by the  viscosity.  It is easy  to  see that the inclusion  of  viscosity
eliminates  the singularity of the solution  in the limit $C_t \to 0$,
$\nu \to 0$.  This is because, after the replacement  indicated
in eq.~(\ref{viscosity}), the wavevector $q  \sqrt{1 - \xi^2}$ in  the
limit $C_t \to 0$ reduces to $\sqrt{-i \omega/\nu}    = \sqrt {\omega /2  
\nu}
(1-i)$ which, for small $\nu$, has a very large positive real  part, and
 therefore
vanishes very rapidly away from the edge.  Correspondingly, a solution
of eq.~(\ref{equationfordispersion})   can always be   found (at small
$q$) if the viscosity is added according to the substitution indicated
in  eq.~(\ref{viscosity}).  In practice, the  viscosity is expected to
be small.   In the  following section   we  shall present our  results
treating  $\nu$, $C_t^2$,  $g$, and  $\omega_c$  as formal parameters.
The actual   values  of these  parameters will   be  discussed  in the
concluding section.

\section {The Dispersion Relation in Various Physical Regimes}

 The behavior  of the  solutions of eq. ~(\ref{equationfordispersion})
as functions of  the dimensionless  parameters $X$  and  $Z$ is rather
complicated.  It is convenient to distinguish three  different regimes 
according to whether the magnetic field,
the edge electric field ($g$), or the shear modulus dominates.
\begin {itemize}
\item {I. Strong Magnetic Field}

This regime is  characterized by  $\vert  q \vert  \ll \vert \omega_c/
\tilde C_t  \vert$  and $\vert  q  \vert \ll \vert \omega_c^2/g\vert$
Therefore $\vert  Z \vert  \gg 1$ and $\vert X \vert \ll \vert Z \vert^2$.  
It is  easy  to  see  that   eq.  ~(\ref{equationfordispersion})  has two
solutions in this limiting case.  The first solution is
 
\begin {equation} \label{longwave1.corr}
(i) ~~~ \xi =  -  {X \over  Z}  + {1 \over  Z}  \left ({X^2  \over  
Z^2} -4
{\sqrt{1 - X^2/Z^2} - 1 \over X^2/Z^2} -4 \right ) \end{equation}  

The explicit form of the dispersion depends on the value of $X$.  The most 
interesting case is $ \vert X \vert \gg \vert Z \vert$ which corresponds to
long wavelengths $\vert q \vert \ll g/C_t^2$ and vanishing shear modulus 
$ \omega_c C_t \ll g$ and is appropriate for the liquid state.  In this case,
 substituting the definitions ~(\ref{parameters}) in eq.~(\ref{longwave1.corr})
 one  obtains, after simple manipulations
\begin {equation} \label {longwave1.XgZ}
\omega  = - {  gq  \over \omega_c} -  {   4C_t^2 q \vert q  \vert\over
\omega_c} + {g^2 \vert q \vert q \over \omega_c^3}  
- 4i \nu {g   \vert  q  \vert^3 \over  \omega_c^2},   
\end{equation}
where we   have used eq.  ~(\ref{viscosity})  to obtain the
imaginary part of the frequency.
 
 Thus, at sufficiently
long  wavelength,   the dispersion  is  linear  and independent of the
visco-elastic constants. 

Recalling the definition
of the  ``gravity  acceleration'' in  eq.~(\ref{defg})   we see that   the
phase velocity for long wavelength $\omega / q$ coincides with
 the classical drift  velocity $v  = cE/B$,  where $E$ is the
magnitude of the electric field at the edge. 
  
Other cases, compatible with our definition of the strong-field regime, 
can also be calculated from eq.~(\ref{longwave1.corr}).  For example, in the 
limit of vanishing electric field $g \to 0$ or large shear modulus, such that 
$\vert X \vert \ll 1$, $\vert  g/C_t^2 \vert  \ll  \vert  q \vert  \ll  \vert
\omega_c /C_t \vert$ we obtain
\begin {equation} \label{medium1}
\omega =  -{gq\over \omega_c}- {2  C_t^2 q|q| \over   \omega_c}  
-4i{C_t^2 \nu  q^4
\over \omega_c^2}. \end {equation}
Setting $g=0$ in this expression, we obtain the dispersion of
  surface elastic waves (``Rayleigh waves'') \cite{Landau7} 
in the presence of a strong 
magnetic field. 
Notice that, due to the  presence of the strong magnetic field,
 the dispersion is quadratic rather than linear in $q$.

The second solution is 
\begin {equation} \label{longwave2.corr}
(ii)~~~~ \xi \simeq Z +(X+4)/Z, \end {equation}  which implies 
\begin {equation} \label {longwave2}
\omega  = \omega_c  sgn(q)+{g q  \over \omega_c}  +  {4 C_t^2q^2 \over
\omega_c}sgn(q)-4i\nu  q^2.   \end  {equation}  

Notice  that   the two
solutions obtained in this section describe waves propagating 
in opposite directions (i.e., the
sign of the  ratio $\omega/q$ is opposite in  the two  cases) and that
the second  solution, for $q\to 0$, is simply  the manifestation of  
the uniform Kohn
mode \cite {Kohn} at the edge of the system.

\item {II. Strong electric field (Gravity Waves)}

Let us now assume that the edge electric field is large, in the sense that
$\omega_c^2/g \ll \vert q \vert \ll g/C_t^2$.  This implies $\vert X \vert \gg 
1$  and $\vert X \vert \gg \vert Z \vert^2$. Notice that $q$ cannot tend 
to zero in this regime, unless $\omega_c =0$. The solutions of 
eq.~(\ref{equationfordispersion}) are then found to be
\begin {equation} \label {longwave3.corr}
\xi \sim \pm \sqrt{X \pm Z \sqrt{X}+4},
\end {equation}
\begin {equation} \label {medium2}
\omega = \pm \sqrt{g \vert q \vert} \left (1 +  {2 C_t^2 \vert q \vert
\over g} \right ) + {\omega_c \over 2}- 2 i \nu q^2 \end {equation}

It is comforting to notice that, for zero magnetic field and shear modulus, 
we recover the classical results for the dispersion and damping of 
``gravity waves'' on the surface of a liquid 
\cite{Landau6}.

\item {III. Large Shear Modulus (Rayleigh Waves)} 

Next, consider the case that the shear modulus dominates, 
in the sense that  $\vert q  \vert  \gg \omega_c/C_t$  and
$\vert q  \vert \gg g /C_t^2$.  This implies $\vert  X \vert  \ll 1$ and
$\vert  Z \vert  \ll 1$.  Again, this definition does not allow one to take the limit $q \to 0$ unless $g$ and $\omega_c$ are both zero.

The doubly  degenerate solution of eq. (\ref{equationfordispersion})  in this
case is $\xi = \pm 0.955$, which implies (for $q\ll \sqrt{C_t^2/\nu}$), 
\begin {equation} \label {shortwave}
\omega =   \pm 0.955 C_t \vert   q  \vert - i    (0.456) \nu q^2. \end
{equation} 

This is  nothing but the familiar  Rayleigh wave on  the surface of an
elastic solid \cite{Landau7},   with   a small  damping due   to    viscosity, 
and no
correction from external electric and magnetic fields. 

\item {IV. Zero Shear Modulus Limit (liquid state)}

In this case (and for zero viscosity) the equations of motion can be
solved exactly.
The expression for the two branches of the dispersion relation is:
\begin{equation}
\omega_\pm=\frac{\omega_c}{2}sgn(q)\pm\sqrt{\Big(\frac{\omega_c}{2}
\Big)^2+g|q|}.
\label{dr}\end{equation}
For  $q\ll \omega_c^2/g$ and $q\gg \omega_c^2/g$
 we recover respectively the expressions (\ref {longwave1.XgZ}) and 
(\ref {medium2}).

The corresponding eigenfunction is given by $u_l$ in eq. (\ref{ul}).

\end {itemize}

We have thus exhausted all  the physically different regimes.  In 
case I to III,  the form of the eigenfunction   is obtained by 
substituting the
appropriate value of $\xi$  in eq. ~(\ref{eigenfunction}).  Results 
for the lower branch of the
 complete  numerical solution  of eq.~  (\ref{equationfordispersion})
exhibiting  crossovers between the  various  regimes discussed in this
section are shown in Fig. 1 and 2. In Fig. 1,  the real part of 
the phase velocity 
$|\omega\omega_c/gq|$ is plotted against  $qC_t/\omega_c$ for decreasing
values of the ratio $R\equiv g/C_t\omega_c$ ($R=1,0.7,0.5$) 
that is for increasing magnetic 
field or shear modulus or decreasing electric field.
The Figure shows clearly the universal long wavelength behavior 
$|\omega\omega_c/gq|\approx 1$ for very small $q$. For $R=0.5$ 
(strong magnetic field
and finite shear modulus)and small $q$   ($q\ll\omega_c/C_t$) the 
exact curve agrees with the 
approximate expression eq. (\ref {medium1}) (labelled in the Fig. as
``lw''). For $R=1$ (strong magnetic field and small shear modulus) 
 the third
term in eq. (\ref{longwave1.XgZ}) becomes predominant and the exact curve lies
below 1. 
For large $q$ the exact curves approach in all cases 
the ``Rayleigh Waves'' regime, 
that is  the lines $|\omega\omega_c/gq|=0.955/R$.

In Fig. 2 we plot the real part of $|\omega\omega_c/2q^2C_t^2|$ versus 
$qC_t^2/g$ for $|q|\ll |\omega_c/C_t|$ for $R=0.01$ (solid line). 
This figure  clearly shows a
 crossover from the linear
dispersion controlled by the electric field to the quadratic dispersion
controlled by the shear modulus  in the
case of strong magnetic field (see eq.(\ref{medium1})).
The dashed line is   the approximate curve for the 
``linear regime'' 
($|q|\ll |g/ C_t^2|$, $|\omega\omega_c/2q^2C_t^2|=|g/2qC_t^2|$), 
while the dotted line corresponds to the ``quadratic
regime''   ($|g/C_t^2|\ll |q|\ll |\omega_c/C_t|$, 
$|\omega\omega_c/2q^2C_t^2|=1$).

In Fig. 3 we plot $|\omega/\omega_c|$ against $qg/\omega_c^2$ for the case
of zero shear modulus (eq. (\ref{dr})). 
For the low frequency branch (labelled with ``$-$''),
we plot also the leading term for long 
wavelength 
($|\omega/\omega_c|=
qg/\omega_c^2$, dashed line)   and  the short wavelength
approximate behavior (eq. (\ref {medium2}), dotted line).

It is interesting to notice that no acceptable solution is found in the
transition region between regimes II and III,  that
is, for $\omega_c/C_t\sinf q\sim g/C_t^2$.

\section{Discussion} Up to this point our classification of different
physical regimes  has  been purely formal: we  have not yet
specified the  values of the  parameters.  We now  wish to   
 state  the concrete 
predictions of our theory for typical systems at  high magnetic 
field.

The value of the electric acceleration at the edge is easily estimated
as 
\begin {equation}
g = {e^2 n_0 \over m} {d \over a}, \label {gest1} \end {equation}
where $n_0 = \rho_0/m$ is the equilibrium density, $d$ is the range of the interaction, and $a$ is the width of the
edge, which is typically of the order of the magnetic length
$l = \sqrt{\hbar c/eB}$.  In a magnetic field, it is convenient  to introduce 
the filling
factor $\nu_0  \equiv 2 \pi n_0 l^2$. Then our estimate for $g$ takes
 the form 
\begin {equation}
g = {e^2 \nu_0 \over 2 \pi \hbar} \omega_c {d \over a}. \label {gest2} 
\end {equation}

The calculation of the shear viscosity is considerably more difficult.
A mode-coupling calculation for  the two-dimensional electron gas {\it
at zero magnetic  field} \cite{Nifosi} yields  numerical results that
can be accurately described by the formula \cite{Conti98} 
\begin{equation} \label{fitnu}
\nu \simeq \left( 59 r_s^{-3/2} + c_1 r_s^{-1}  + c_2 r_s^{-2/3} + c_3
 r_s^{-1/3}\right)^{-1} {\hbar  \over m}, 
\end{equation}
 where   $r_s  = \sqrt{1/\pi  n_0  a_0^2}$   is the usual  electron gas
parameter, $a_0 = \hbar^2/m e^2$ is the effective
Bohr radius of  the host semiconductor,  and $c_0 = 0.25$, $c_1 =20.6$,
$c_2=22.7$, $c_3=12.8$.  We are  not aware  of any calculation  of
the viscosity  in the  presence of  a  magnetic  field, but we  expect
eq.~(\ref{fitnu}) to give  at least the right  order of magnitude at a
given density.

As for the shear modulus, we expect it to vanish, if the system is liquid,
 at low frequency, leaving us with $\tilde  C_t^2 = -i \omega
\nu$.   If, instead, the system is a solid,  then $C_t^2$ has a finite
value which can be estimated from dimensional considerations:
\begin {equation} \label{shearmodulus}
C_t^2 \simeq { e^2 n_0^{1/2} \over m} 
\end {equation} It is the viscosity that can be
neglected in this case. 
  For example, in the classical
two-dimensional  Wigner crystal with Coulomb interactions ($d = \infty$), the exact result 
is  $C_t^2   \simeq 0.24    e^2 n_0^{1/2}/m$ \cite{Conti98}.

Let us first consider the dispersion of edge waves in a {\it liquid}. 
Because $C_t = 0$, and $\nu$ is very small 
 we immediately see that $\vert X \vert \gg 1$ and
$\vert Z \vert \gg 1$ at all realistic wavevectors $q \ll 1/a \sim 1/d$. 
The inverse length $\omega_c^2 /g$ is given by

\begin {equation} \label{omegac2g}
{\omega_c^2 \over g} = a_0^{-1} { 2 \pi \over \nu_0} { a \over d} \left ( {a_0 \over l} \right )^2.  \end {equation}
This is of the order of $1/a$ or larger for typical quantum Hall systems in which the density is $n_0 \sim 10^{10} - 10^{11} cm^{-2}$ and the filling factor is less than $1$.  Therefore, these systems fall within the ``strong magnetic field regime'' of Section IV. 
 Because the frequency
vanishes linearly with $q$ the ratio $X/Z = g/\omega_c \sqrt{-i \omega
\nu}$ tends to  infinity,  and the dispersion  is therefore   given by
eq.~(\ref{longwave1.XgZ}) 
\begin {equation}
\omega  = -{gq \over  \omega_c} + {g^2q|q| \over  \omega_c^3}
- 4  i \nu  {g  \vert q \vert^3 \over
\omega_c^2}. \label{fluiddispersion}\end {equation}  The eigenfunction
has the simple form 
\begin {equation}
{\bf u} \propto [{\bf \hat x} + i sgn (q) {\bf \hat y}] e^{\vert q \vert
x +iqy}, 
\label {highfieldeigenfunction} \end {equation} 
which describes a circular motion of each volume element. 
Then, making use of our estimate ~(\ref{gest2}) for the electric acceleration
 the phase velocity of the wave $v = g/\omega_c = (e^2 \nu_0/h) (d/a)$ is 
obtained at once.

Let us now consider the case that the low-frequency shear modulus does
{\it  not} vanish:  this would happen,   for example, if the electrons
solidified in a  Wigner crystal structure.   According  to our general
discussion it might be  possible, with increasing wavevector, to cross
over to  an    ``intermediate wavelength regime  ",   in   which the
dispersion is controlled   by the shear modulus,  and  varies as $q^2$
(see eq.~(\ref{medium1})). 

Unfortunately, this crossover is not likely to occur within the region
of wavevectors in which  our  theory applies,  namely $q <<  1/a \sim 1/d$.
Indeed,  from eqs.~(\ref{gest1}), ~(\ref{gest2}),  and
~(\ref{shearmodulus})
we see that, for $d$ comparable to, but somewhat larger than  the
average distance between electrons, both  the ratios  $g/C_t^2$,
and $\omega_c /C_t $ are of the order of the inverse of the inter-electron
distance.   This implies that for $q  \ll 1/a, 1/d$ both
$X$ and
$Z$ are $\gg 1$, and  we  clearly fall into  the ``long wavelength regime"
of the  previous  section.  The only possibility  to  observe the crossover 
to
shear-modulus-sustained  waves in the present model,  would arise if
the range of the interaction $d$ were much less than  the typical 
inter-electron
distance -- not an easily realizable situation.

Thus, in summary, we have  shown that the two-dimensional electron gas
on a   neutralizing  background of   charge sustains  only   {\it one}
macroscopically charged collective mode  which decays exponentially as
one moves away  from the edge. {\it The scale  of this exponential decay is
the same as  the scale of variation of  the density along the edge}. We
have derived analytical  and numerical expressions for the  dispersion
relations and the  eigenfunctions   in  various  physically   distinct
regimes.  An important result  of our study  is that the  character of
the edge excitations in this model is controlled almost exclusively by
the strength of  the electric field at the  edge, and does not  depend
significantly, on the shear modulus: therefore,  the dispersion is the
same, to leading order in $q$, for the liquid and the solid state (see
eq.~(\ref {longwave1.corr})). 

 These  results  suggest that  the   behavior of   the $I-V$  tunneling
characteristics at low bias voltage would be the same at the edge of
a liquid and of a Wigner crystal: in particular, the power law $I \sim
V^{1/\nu_0}$, where $\nu_0$ is the bulk filling factor, is expected in
both cases \cite{Conti1}. 

\section {Acknowledgements} This work was supported by NSF Grant No. 
DMR-9706788. G.V.   gratefully  acknowledges  the hospitality  of  the
Institute for Theoretical  Physics  at the University   of California,
Santa Barbara, under NSF Grant No. PHY94-07194, were part of this work
was done.  We thank M. Fogler for several interesting discussions. 

\appendix
\section{Extension to long-range interaction} 

In this  paper  we     have taken  advantage  of    some  mathematical
simplifications arising from the assumption that the electron-electron
interaction is of  finite  range in  space.  Extending  the theory  to
properly include Coulomb interactions is nontrivial.  In this Appendix
we want  to sketch an approximate method  to do this  extension, which
entails minimal changes in the structure of the equations.  The method
is non rigorous, yet  it yields qualitatively  correct  results for the
long wavelength dispersion. 

To  begin,   we  observe  that,   in the   equation   of motion
~(\ref{equationofmotion}) the term $C^2{\bf \nabla}({\bf \nabla} \cdot
{\bf u})$ should   be replaced (ignoring  the  small contribution from
$\tilde K$) by 
\begin {equation} \label{longrangeterm}
{\bf \nabla} \int   {e^2 \over \vert {\bf  r}  - {\bf r'} \vert}  {\bf
\nabla '} \cdot {\bf u} ({\bf r'}) d {\bf r'}. 
\end {equation}

In an  infinite system  this  term would pose  no  problem: by Fourier
transformation   it   could      be    recast   in   the      form  of
eq.~(\ref{equationofmotion}) with a $q$-dependent longitudinal ``sound
velocity''
  
\begin {equation} 
\label{clq}
 C^2 ({\bf q}) = {2 \pi  e^2 n_0 \over m q} 
\end {equation}
 which tends to infinity
as $q$  tends to zero. In  a semi-infinite system, however, things are
not so simple, and the  Coulomb interaction cannot be simply  absorbed
in a $q$-dependent sound velocity. 

The idea of  our approximation scheme is to  neglect the effect of the
edge on  the bulk equation of motion,  which therefore retains the form
of  eq. ~(\ref{equationofmotion}), with  $C^2$  tending to infinity in
the long wavelength limit.  The effect of  the edge will be taken into
account only via the boundary conditions ~(\ref{bc1}) and (\ref{bc2}),
which force the solution to be a certain superposition of bulk waves. 

So   far  the   theory is    formally   identical  to the  short-range
case. However,  observe that the   key quantity $g$, which enters  the
boundary conditions, is   ill-defined  in  the  case of  the   Coulomb
interaction.    The external potential  in  this  case  is simply  the
electrostatic potential created  by a uniform distribution of positive
charge precisely compensating the electronic charge in the half-plane.
The electric  field    produced  by  this charge  tends    to infinity
(logarithmically) at the edge of the half-plane,  and therefore $g$ is
infinite (see eq.~(\ref{defg})). 

In order to obtain the correct form of  the boundary conditions in the
Coulomb  case   we return   to  eq.~(\ref{bc00})  and   note that
  from the
equilibrium condition $\sigma^{(0)}_{xx} ({\bf  r})$ is given by 
\begin {equation} \label {equilibriumcoulombstress}
\sigma^{(0)}_{xx} ({\bf  r}) = -  n_0^2  \int\int_{x'< 0}
{e^2 \over  \vert  {\bf r} - {\bf  r'}  \vert}d  {\bf  r'}, \end
{equation} where the integral is restricted to the half-plane $x'<0$. 

Consider a point in the vicinity of the geometric edge with 
${\bf r} = (\epsilon+u_x(0,y),y)$, where $\epsilon$ is a length of
the order of the physical width of the edge $a$. Similarly to
what we did in the short range case in sec. II we now expand 
$\sigma^{(0)}_{xx}$ to first order in $u_x$ and discard the constant
zero order contribution. After integrating with respect to $x'$,
 we obtain 
\begin {equation} \label {linearcoulombstress1}
\sigma^{(0)}_{xx} (u_x,y)  \simeq  n_0^2 
\int_{-  \infty}^{\infty} {e^2 \over \sqrt{\epsilon^2+(y-y')^2}}u_x(0,y') dy'
. 
\end{equation}

Specializing to solutions   with definite  wavevector along the   edge
(i.e.,    $u_x(0,y)    \propto     e^{iqy}$)       we     see     that
eq.~(\ref{linearcoulombstress1}) takes the simpler form 
\begin {equation} \label {linearcoulombstres21}
\sigma^{(0)}_{xx} (u_x,y) \simeq  n_0^2  2
e^2K_0(\vert q\epsilon \vert) u_x(0,y),  \end  {equation} 
where $K_0$ is  the
modified Bessel function.  Unfortunately, the $\epsilon \to  0$ limit of this
expression does  not exist, due to the  logarithmic divergence  of the
Bessel function  $K_0(x) \sim -  ln(x)$ for  $x  \to 0$.  However, the
divergence is very weak, and, in view of the fact that the position of
the edge is defined within  an uncertainty of  the order of $a$ ($\sim
0$ in  our theory)  it is  legitimate  (with logarithmic  accuracy) to
replace $ K_0(\vert q \epsilon\vert)$ by $K_0(\vert qa \vert)$.

Substitution  of this expansion in  eq.~(\ref{bc00})  leads to the new
boundary conditions:  these can still be written  in the  form of eqs.
~(\ref{bcq}), but now   both  the longitudinal  sound
velocity, and the electrical acceleration  $g$ are functions of $q$: the
former is given by eq.~(\ref{clq}), while for the latter 
\begin {equation} \label{geff}
g  \to g(q)  =  2 e^2  {n_0 \over m}K_0(\vert qa
\vert). 
\end {equation}

From this point on, all the calculations proceed as in the short-range
 case.  In particular, the logarithmic divergence of the effective $g$
 causes the dispersion of the edge waves to vary as $-(2\nu_0 e^2/h) q
 \ln(qa)$ in a   magnetic field, and   as  $\sqrt{2e^2n_0}
\sqrt{q  ln(qa)}$   without a
 magnetic field. 

It  is amusing  to observe  that   in the three-dimensional case   the
effective $g$ at the surface would be $g(q) =  2 \pi \rho_0 e^2/m^2 q$
, leading,  in  the absence of a  magnetic  field, to the  well  known
result for the frequency    of  surface  plasmons,  namely  $\omega   =
\sqrt{gq} = \sqrt{2 \pi e^2 n_0/m}$.

\begin{figure}
\caption{
Real part of the dimensionless phase velocity $|\omega\omega_c/gq|$ versus
$qC_t/\omega_c$
for three values of $R=g/C_t\omega_c$.
For 
each curve the short wavelength limit $|\omega\omega_c/gq|
=0.955/R$ - coinciding with the dispersion of classical Rayleigh waves - 
is plotted for comparison. For the curve with $R=0.5$ we 
also plot the
long wavelength limit  $|\omega\omega_c/gq|=1+2C_t^2|q|/g$ 
 (``lw'' curve). 
}
\label{Fig1}
\end{figure}

\begin{figure}
\caption{
Real part of $|\omega\omega_c/2q^2C_t^2|$ versus 
$qC_t^2/g$ for $R=0.01$ (solid line).
The dashed line is   the approximate curve for the 
``linear regime'' 
($|q|\ll |g/ C_t^2|$, $|\omega\omega_c/2q^2C_t^2|=|g/2qC_t^2|$), 
while the dotted line corresponds to the ``quadratic
regime''  ($|g/C_t^2|\ll |q|\ll |\omega_c/C_t|$, 
$|\omega\omega_c/2q^2C_t^2|=1$).
}
\label{Fig2}
\end{figure}
\begin{figure}
\caption{
Real part of the upper (``+'') and lower (``$-$'') branch of
the dimensionless dispersion curve $|\omega/\omega_c|$
versus  $qg/\omega_c^2$ for vanishing shear modulus.
The dashed line represents the long wavelength limit 
($|\omega/\omega_c|=qg/\omega_c^2$) and the dotted line  
the short wavelength limit ($|\omega/\omega_c|=\sqrt{qg/\omega_c^2}-1/2$)
of the lower branch.
}
\label{Fig3}
\end{figure}

\newpage
\unitlength1cm 

\begin{picture}(20,20.0)
\put(-8.5,-13.0){
\makebox(10,15){
\includegraphics{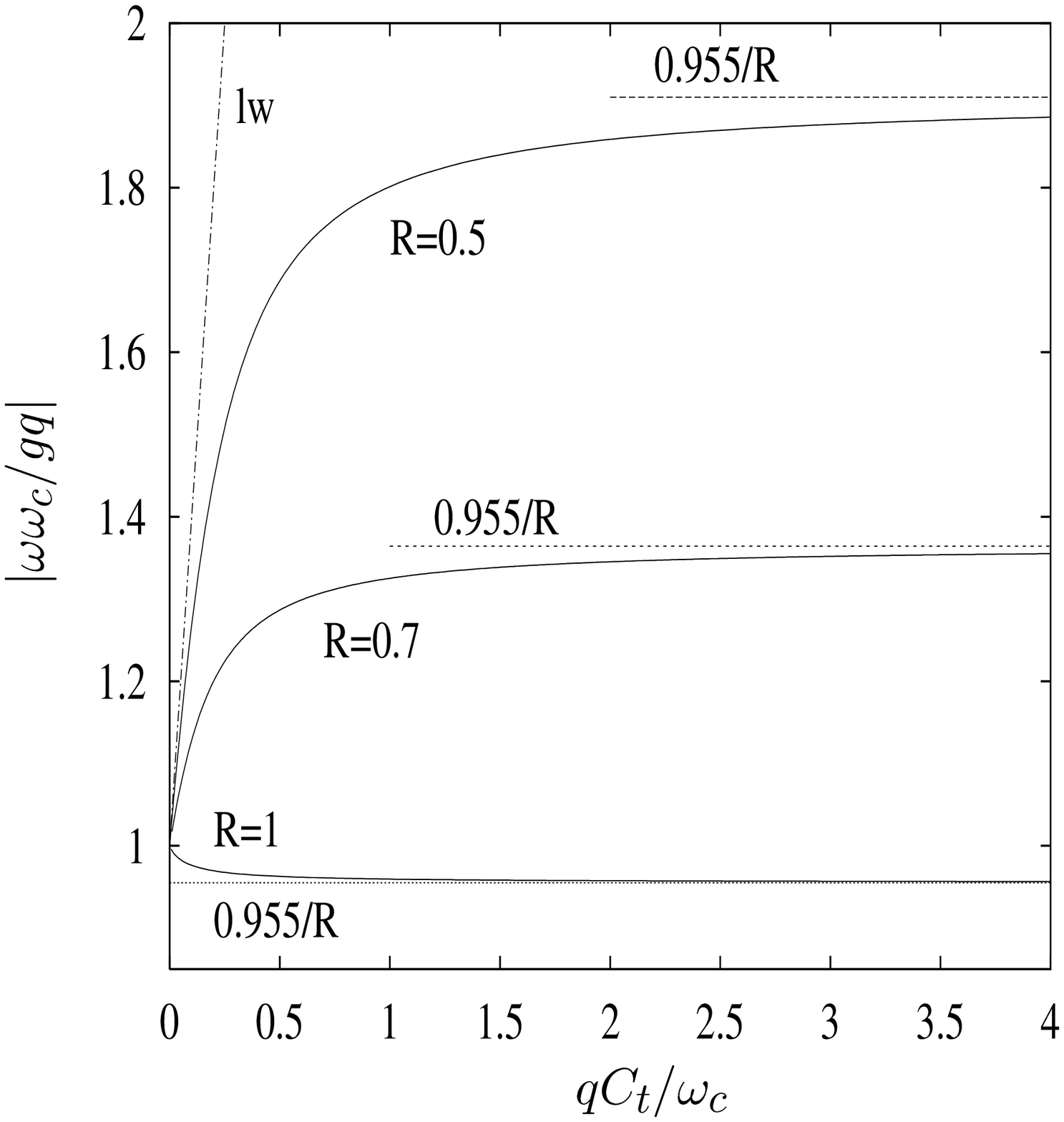}
}
}
\put(0,-2.5){\LARGE Figure 1 }
\end{picture}
\newpage
\unitlength1cm

\begin{picture}(20,20.0)
\put(-8.5,-13.0){
\makebox(10,15){
\includegraphics{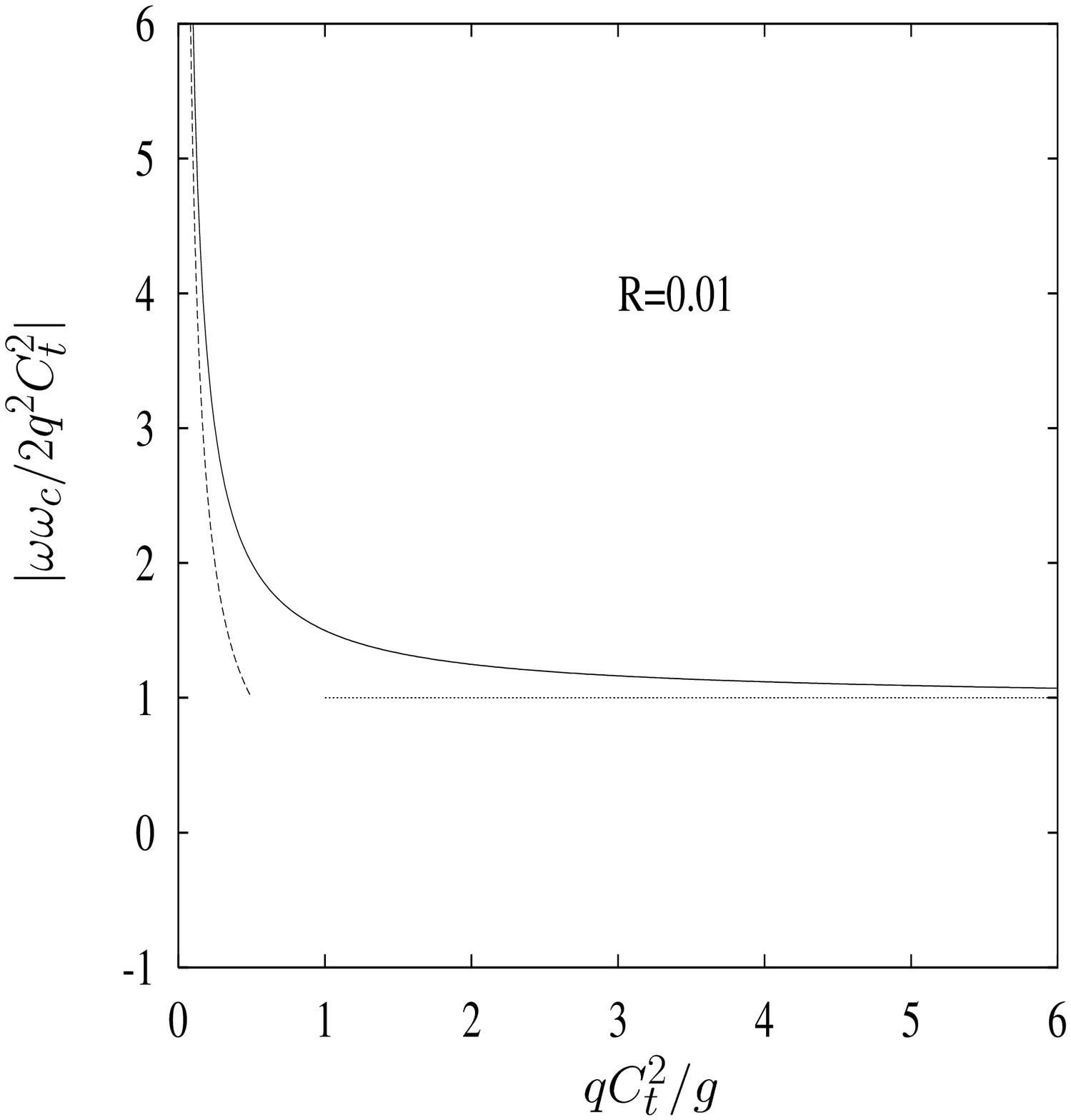}
}
}
\put(0,-2.5){\LARGE Figure 2 }
\end{picture}

\newpage
\unitlength1cm
\begin{picture}(20,20.0)
\put(-8.5,-13.0){
\makebox(10,15){
\includegraphics{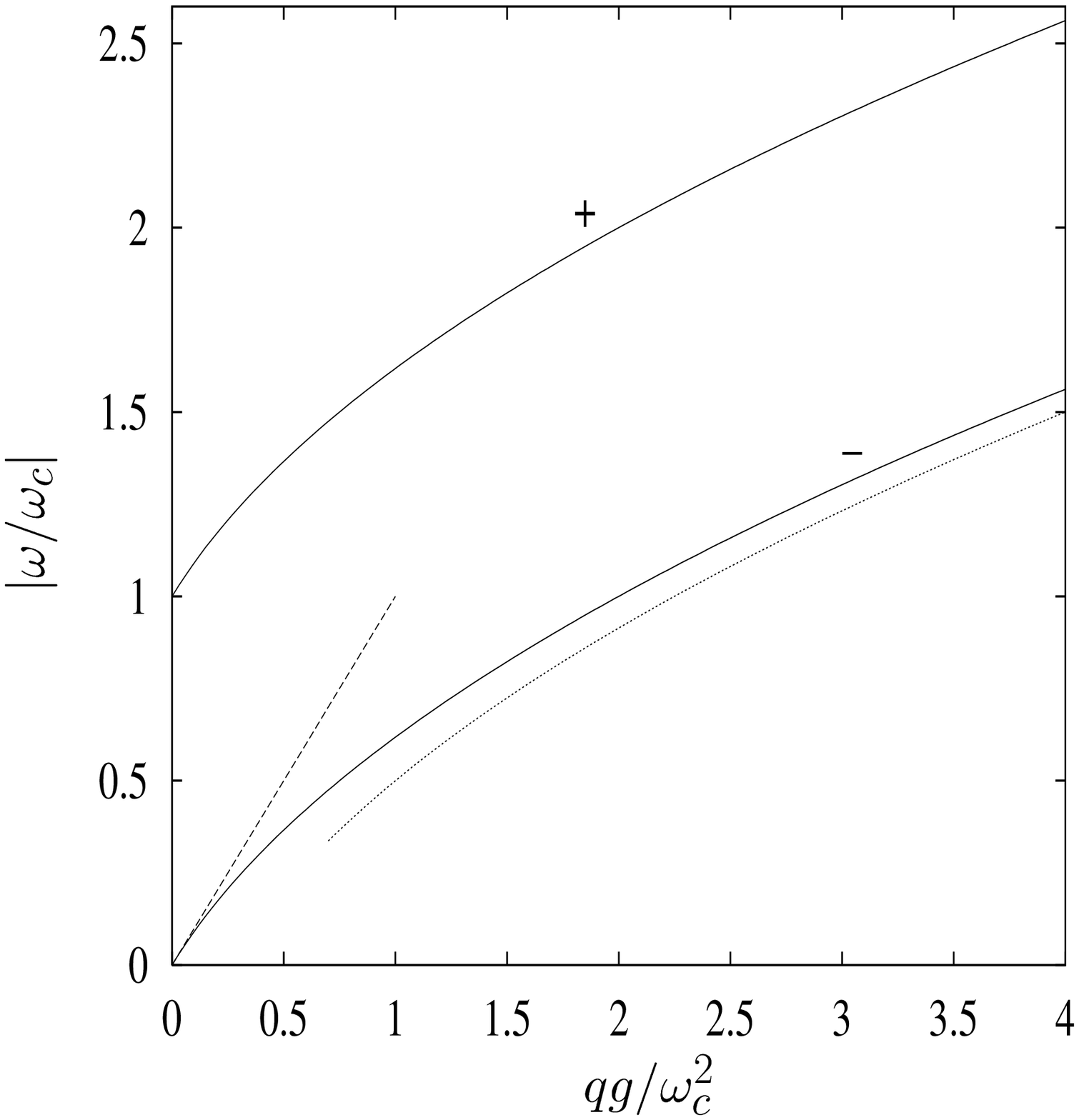}
}
}
\put(0,-2.5){\LARGE Figure 3 }
\end{picture}

\end{document}